\documentclass[12pt]{article}
\newif\ifpdf
\ifx\pdfoutput\undefined
\pdffalse 
\else
\pdfoutput=1 
\pdftrue
\fi

\ifpdf
\usepackage[pdftex]{graphicx}
\else
\usepackage{graphicx}
\fi

\usepackage{ae,amsmath,amssymb,exscale,subfigure,natbib}

\DeclareMathOperator{\diag}{diag}

\makeatletter

\long\def\@makecaption#1#2{%
\scriptsize%
 \vskip\abovecaptionskip
  \sbox\@tempboxa{#1. #2}%
  \ifdim \wd\@tempboxa >\hsize
   #1. #2\par
  \else
    \global \@minipagefalse
    \hb@xt@\hsize{\hfil\box\@tempboxa\hfil}%
  \fi
  \vskip\belowcaptionskip}
\makeatother

\date{}

\begin{document}
\title{{\bf An ordinary differential equation model \\ for the multistep transformation to cancer}}
\author{Sabrina L. Spencer$^{a,b,*}$, Matthew J. Berryman$^{b}$, \\Jos\'{e} A.
Garc\'{i}a$^{c}$, and Derek Abbott$^{b}$} \ifpdf
\DeclareGraphicsExtensions{.pdf, .jpg, .tif} \else
\DeclareGraphicsExtensions{.eps, .jpg} \fi \maketitle
\footnotesize $^{a}$Department of Human Genetics, University of
Michigan, Ann Arbor, MI 48109-0618, USA. $^{b}$Center for
Biomedical Engineering and School of Electrical \& Electronic
Engineering, The University of Adelaide, SA 5005, Australia.
$^{c}$Laboratory of Theoretical Biology, Universidad La Salle,
06140 M\'{e}xico, D.F., M\'{e}xico.

* Corresponding author.  Email:  sabrinal@umich.edu, Telephone:
+1-856-327-5283

\begin{abstract}
Cancer is viewed as a multistep process whereby a normal cell is
transformed into a cancer cell through the acquisition of
mutations.  We reduce the complexities of cancer progression to a
simple set of underlying rules that govern the transformation of
normal cells to malignant cells. In doing so, we derive an
ordinary differential equation model that explores how the balance
of angiogenesis, cell death rates, genetic instability, and
replication rates give rise to different kinetics in the
development of cancer. The key predictions of the model are that
cancer develops fastest through a particular ordering of mutations
and that mutations in genes that maintain genomic integrity would
be the most deleterious type of mutations to inherit. In addition,
we perform a sensitivity analysis on the parameters included in
the model to determine the probable contribution of each. This
paper presents a novel approach to viewing the genetic basis of
cancer from a systems biology perspective and provides the
groundwork for other models that can be directly tied to clinical
and molecular data.

{\it Keywords}:  cancer; oncogenesis; genetic instability;
multistep transformation; ordinary differential equation model
\end{abstract}

\section{Introduction}
The standard perspective on cancer progression is that it is a
form of somatic evolution where certain mutations give one cell a
selective growth advantage~\citep{cahill99}. Oncogenesis is thought
to require several independent, rare mutation events to occur in
the lineage of one cell~\citep{nowell}. Kinetic analyses have shown
that four to six rate-limiting stochastic mutational events are
required for the formation of a tumor ~\citep{Armitage,Renan}.
\cite{hanahan00} proposed the following six
hallmark capabilities that a normal cell must acquire to become a
cancer cell: (i) self-sufficiency in growth signals, (ii)
insensitivity to anti-growth signals, (iii) evasion of apoptosis,
(iv) limitless replicative potential, (v) sustained angiogenesis,
and (vi) tissue invasion and metastasis. They define genetic
instability as an ``enabling characteristic''  that facilitates
the acquisition of other mutations due to defects in DNA repair
processes. We reduce these characteristics to the following four:
angiogenesis ($A$), immortality, including evasion of cell death
($D$), genetic instability, a function of mutation rates ($G$),
and increased replication rate ($R$). We consider invasion and
metastasis ($M$) as a final step that allows the spread of a
localized tumor.  In line with the views of \cite{hanahan00}, we foresee cancer research developing into a
logical science where the molecular and clinical complexities of
the disease will be
 understood in terms of a few underlying principles. We
therefore explore the multistep progression to cancer with an
ordinary differential equation (ODE) model which, despite the
apparent complexity of the equations, is based on basic principles
and a minimal set of parameters.
\section{Structure and parameters of the model}
Although the model applies to the process of oncogenesis in
general, the parameters are loosely based on breast cancer data.
We consider the following cell populations: a population of
$10^{8}$ normal cells ($N$), cells which have acquired the ability
to induce angiogenesis ($A$), cells with mutations which allow
them to avoid death ($D$), cells with mutations that lead to
genetic instability ($G$), cells with mutations which increase
their replication rate ($R$), and cells with two or more of these
mutations. Cell populations that have acquired two or three
mutations are denoted by listing the mutations together in
alphabetical order (state $DRA$ would be listed as state $ADR$,
for example).  We label a cell which has acquired all four
mutations a primary tumor cell ($T$). Although our model only
addresses development of a primary tumor in genetic detail, we
allow a primary tumor cell that has acquired the capability to
invade and metastasize to become a metastatic cell ($M$).

The spontaneous mutation rate in human cells has been estimated to
be in the range of $10^{-7}$ to $10^{-6}$ mutations/gene/cell
division~\citep{jackson98}. We assume a spontaneous mutation rate
of $k_{1}=10^{-7}$ mutations/gene/cell division. The loss of DNA
repair genes can increase the mutation rate by a factor ranging
from $10^{1}$ to $10^{4}$~\citep{Tomlinson1996}.  We assume that
the mutation rate after a genetic instability mutation increases
1000-fold to $k_{2}=10^{-4}$ mutations/gene/cell division.
Successful invasion and metastasis depend upon acquisition of the
other hallmark capabilities, as well as several new
capabilities~\citep{hanahan00}.  To simplify the model, we do not
address the multistep progression of a tumor cell to a metastatic
cell, and instead consider this complex process as one step. It
has been estimated that the rate of successful metastasis is in
the range of $10^{-9}$ to $10^{-7}$ per cell division
~\citep{mool}, and so we use a conservative estimate of
$k_{3}=10^{-9}$ for the transition from a primary tumor cell to a
metastatic cell.

A tumor cannot grow past about $10^{6}$ cells without angiogenesis
supplying blood to the tumor~\citep{Folkman}.  We thus cap the
size of the tumor at $10^{6}$ cells until at least 10\% of the
population of non-normal, non-metastatic cells has acquired a
mutation in an $A$ gene. This accounts for the fact that only a
fraction of the cells in a tumor need to send angiogenesis signals
in order to develop an adequate blood supply for the tumor.  In
addition, populations of non-normal, non-metastatic cells are
always capped by a lethal tumor burden limit of $10^{13}$
cells~\citep{surgonc}, irrespective of the angiogenesis cap.

A survey of microarray data published on number of genes up- or
down-regulated in various
cancers~\citep{genes1,genes2,genes3,genes4} led us to estimate that
there are approximately $400$ genes involved in development of a
primary tumor.
 Dividing by four categories, we assume that there are approximately 100 genes involved in
each of categories $A$, $D$, $G$, and $R$. To account for the fact
that some genes function in more than one category, we allow
double and triple state transitions but reduce the number of genes
involved to the order of 10 and 1, respectively, to reflect the
likelihood that a single mutation would affect more than one
category. For example, we assume there are 100 genes involved
transitions where only one mutation is acquired (e.g.
$N\rightarrow A$, $D \rightarrow DR$, $AG \rightarrow AGR$, or
$ADG \rightarrow ADGR$), 10 genes involved in transitions where
two mutations are acquired in one step (e.g. $N \rightarrow AD$,
$G\rightarrow ADG$, or $AR \rightarrow ADGR$), and 1 gene
involved in transitions where three mutations are acquired in one
step (e.g. $N \rightarrow ADG$ or $G \rightarrow ADGR$). This
feature accounts for a mutational hit in p53, for example, which
could take a cell directly from N to DGR, as p53 is involved in
apoptosis, DNA repair, and cell cycle progression~\citep{surfing}.

We estimate that the relative contribution to increased net
proliferation for mutations in the $D$ and $R$ categories is $0.7$
and $0.3$, respectively, an inference made from work by \cite{tomlinson2}. Using this $D$:$R$ ratio of 0.7:0.3,
a tumor volume doubling time for breast cancer of 500
days~\citep{surgonc}, and a cell division rate for breast cancer of
$1/10.00$ days$^{-1}$ ~\citep{divisions}, we calculate (see section
~\ref{cell div and cell death rates}) the following: cells without
a mutation in an $R$ gene divide every $b=10.00$ days, cells with
a mutation in an $R$ gene divide every $b_{R}= 9.89$ days, the
lifetime of cells without a mutation in a $D$ gene is $d=10.00$
days, and the lifetime of cells with a mutation in a $D$ gene is
$d_{D} = 10.16$ days. The birth and death rates are equal for
normal cells and for all cells without a mutation in $D$ or $R$.

\begin{figure}[phtb]
\centering
\includegraphics[scale=0.65]{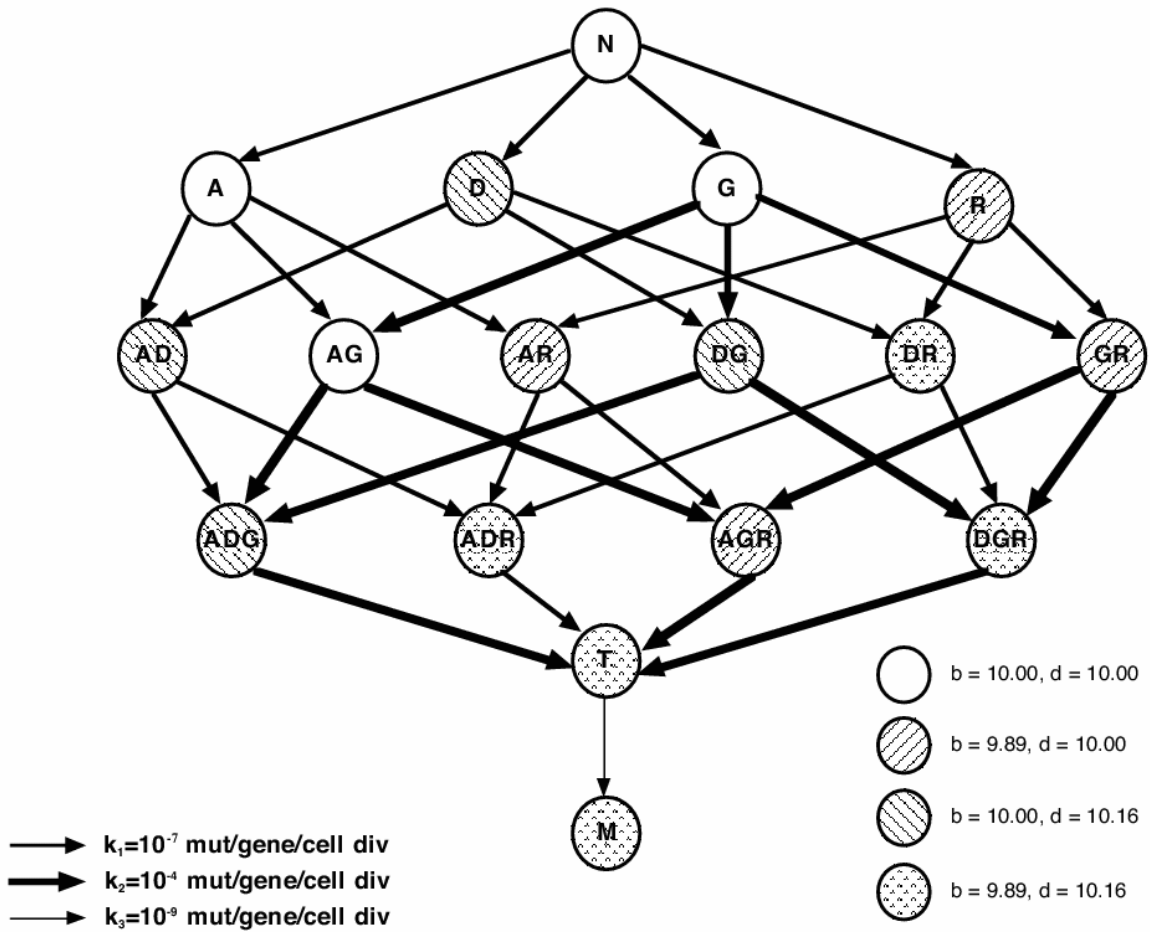}
\caption{State diagram of the model. Normal cells $(N)$ can
acquire mutations which give the cell the capability to induce
angiogenesis $(A)$, mutations which give the cell the capability
to avoid death $(D)$, mutations which lead to genetic instability
$(G)$, or mutations which increase the proliferation rate $(R)$.
These mutations are acquired at rate $k_{1}$. After a mutation in
$G$, the mutation rate increases to $k_{2}$. Cells with one
mutation go on to acquire two, three, and four mutations, denoted
by listing the mutations together in alphabetical order for the
cases of two and three mutations.
 When a cell has acquired all four mutations, it becomes a
primary tumor cell $(T)$. Finally, tumor cells become metastatic
cells $(M)$ at rate $k_{3}$. Double and triple state transitions
are also allowed, as detailed in the text, but are not shown in
this diagram for simplification.  Cell birth rates ($1/b$) and
cell death rates ($1/d$) have units days$^{-1}$.} \label{model}
\end{figure}

The above information is depicted in a unified fashion in
Figure~\ref{model} and the parameters appearing in the ODE model
are given in Table~\ref{valtable}.
\begin{table}[htb]
\scriptsize \centering \caption{Parameters appearing in the ODE
model.} \hspace*{\fill}\newline
\begin{tabular}{|c|c|c|c|}
\hline Characteristic & Parameter & Value & Reference
\\\hline\hline Mutation rate without a $G$ mutation & $k_{1}$ & $10^{-7}$ mut./gene/cell div. & ~\citep{jackson98}
\\\hline Mutation rate with a $G$ mutation & $k_{2}$ & $10^{-4}$ mut./gene/cell div. & ~\citep{Tomlinson1996}
\\\hline Metastasis rate & $k_{3}$ & $10^{-9}$ /cell division & ~\citep{mool}
\\\hline Cell division rate without an $R$ mutation & $1/b$ & $1/10.00$ days$^{-1}$ & ~\citep{divisions}
\\\hline Cell division rate with an $R$ mutation & $1/b_{R}$ & $1/9.89$ days$^{-1}$ & see section ~\ref{cell div and cell death rates}
\\\hline Cell death rate without a $D$ mutation & $1/d$ & $1/10.00$ days$^{-1}$ & ~\citep{divisions}
\\\hline Cell death rate with a $D$ mutation & $1/d_{D}$ & $1/10.16$ days$^{-1}$ & see section ~\ref{cell div and cell death rates}
\\\hline Cap when $\leq10\%$ of cells have $A$ & & $10^{6}$ cells & ~\citep{Folkman}
\\\hline Lethal tumor burden cap on non-$N$, non-$M$ cells  & & $10^{13}$ cells & ~\citep{surgonc}
\\\hline
\end{tabular}
\label{valtable}
\end{table}

In the next section, we present an ODE model that will be used to
explore the following areas:
\begin{enumerate}
\item The kinetics of various paths to cancer. \item The effect of
inherited mutations on cancer development. \item A sensitivity
analysis of variations in the parameters.
\end{enumerate}

\section{Construction of the ODE model}

Based on the basic rules outlined in the state diagram in
Figure~\ref{model}, we construct 17 ODEs to model a heterogeneous
population of cells undergoing the multistep process of
tumorigenesis.  Each equation represents one of the 17 populations
of cells depicted in the state diagram and has the following
format: the population of cells in a state is increased by cells
gaining mutations and entering that state from previous states, is
increased by cells replicating and remaining in that state, and is
decreased by cells gaining new mutations and leaving that state
for a new state.  The populations are capped by two logistic
terms, as detailed below.

We condense the ODEs into vector format as follows:

\begin{equation}
\frac{d\mathbf{y}}{dt} =
\begin{cases}\left(
\diag\left( \diag\left(\mathbf{y}^{T}\mathbf{k}
\right)^{T}\mathbf{b}\right) \mathbf{M}+
\diag\left(\left(\mathbf{b}-\mathbf{d}\right)^{T}\mathbf{y}\right)
\right)  \mathbf{S}
\left(1-\frac{P_{\overline{NM}}}{10^{13}}\right) + \mathbf{m_{m}},
 & \text{no angiogenesis cap},\\
 \left(
\diag\left( \diag\left(\mathbf{y}^{T}\mathbf{k}
\right)^{T}\mathbf{b}\right) \mathbf{M}+
\diag\left(\left(\mathbf{b}-\mathbf{d}\right)^{T}\mathbf{y}\right)
\right)  \mathbf{S}
\left(1-\frac{P_{\overline{NM}}}{10^{13}}\right)\\\times
\left(1-\frac{P_{\overline{NM}}}{10^{6}}\right) + \mathbf{m_{m}},
& \text{angiogenesis cap},
 \end{cases}
 \label{ODEs}
\end{equation}
where $\mathbf{y}$ is the row vector of cell populations; $y_{1}$
is the population of normal cells, $y_{2},y_{3},\ldots,y_{15}$ are
the populations of cells with single, double, and triple
mutations, $y_{16}$ is the number of primary tumor cells (cells
with all four mutations), and $y_{17}$ is the number of metastatic
cells. Here, $\diag\left(\cdot\right)$ is the operator which forms
the row vector of the main diagonal of the matrix. The
corresponding rate (row) vector is $\mathbf{k}$, with mutation
rates $k_{i}$ (mutations/gene/cell division) corresponding to the
mutation rate for element $y_{i}$ in $\mathbf{y}$. The same
applies to the birth
 rates $\mathbf{b}$ (day$^{-1}$) and death rates $\mathbf{d}$ (day$^{-1}$). The metastasis rate vector is
 $\mathbf{m_{m}}=\left(0, 0, \ldots, 0, 10^{-9}\times y_{16}\right)+(1/b_{R}-1/d_{D})y_{17}$, corresponding to cells leaving $y_{16}$ for $y_{17}$
 at rate $10^{-9}$, and a doubling of metastatic cells at rate $(1/b_{R}-1/d_{D})$ for $1/b_{R}$ and $1/d_{D}$ as given in Table~\ref{valtable}.
 The $17\times 17$ upper triangular matrix $\mathbf{M}$ consists of elements $M_{i,j}$
($j\neq i$) for the number of genes associated with going from
state $i$ to state $j$, and
\begin{equation}
M_{i,i}=- \displaystyle \sum_{j\neq i}M_{i,j},
 \label{M}
\end{equation}
is the main diagonal containing the number of genes for leaving each of the states.
$\mathbf{S}$ is the $17\times17$ matrix
\begin{equation}
\mathbf{S}=\begin{pmatrix}
0 & 0 & 0 & \ldots & 0\\
0 & 1 & 0 & \ldots & 0\\
\vdots &  & \ddots & & \vdots\\
0 & \ldots  & & 1 & 0\\
0 & \ldots  & & 0 & 0
\end{pmatrix}
\label{S}
\end{equation}
used to apply the cell population caps to the non-normal,
non-metastatic cells.  Non-normal, non-metastatic cells are
denoted by $P_{\overline{NM}}$, where
\begin{equation}
P_{\overline{NM}}=\left(\sum_{i=2}^{16}y_{i}\right).
\label{capterm}
\end{equation}
The system is capped at $10^{6}$ cells using a logistic term if
$<10\%$ of the non-normal, non-metastatic cells are in states with
angiogenesis mutations, otherwise this term is removed.   The
populations of non-normal, non-metastatic cells are also capped by
a lethal tumor burden limit of $10^{13}$ cells~\citep{surgonc},
irrespective of the angiogenesis cap. The ODEs are solved using the Runge-Kutta method of order 5, with a variable step size between $1$ and $10^{-5}$, to guarantee the errors in calculating the populations remain within $10^{-4}$.

Below, we have reproduced the normal cell and four single state
ODEs from the compact vector form for ease of comprehension. Note
that our equations assume a constant, renewing population of
normal cells, since we assume that cells leaving state $N$ for
other states are few enough in number so as not to affect the
population of $N$ cells.
\begin{subequations}
\begin{align}
\frac{dP_{N}}{dt}&=0,
 \label{N}\\
\frac{dP_{A}}{dt}&=\left(\frac{100P_{N}k_{1}}{b}-\frac{\left(3\times100+
3\times10+1 \right)P_{A}k_{1}}{b}
\right)\left(1-\frac{P_{\overline{NM}}}{10^{6}}\right)\left(1-\frac{P_{\overline{NM}}}{10^{13}}\right),
\label{A}\\
\frac{dP_{D}}{dt}&=\left(\frac{100P_{N}k_{1}}{b}+P_{D}\left(\frac{1}{b}-
\frac{1}{d_{D}}\right)-
\frac{\left(3\times100+3\times10+1\right)P_{D}k_{1}}{b}\right)
\left(1-\frac{P_{\overline{NM}}}{10^{6}}\right)\left(1-\frac{P_{\overline{NM}}}{10^{13}}\right),
\label{D}\\
 \frac{dP_{G}}{dt}&=\left(\frac{100P_{N}k_{1}}{b}-
\frac{\left(3\times100+3\times10+1\right)P_{G}k_{2}}{b}\right)\left(1-
\frac{P_{\overline{NM}}}{10^{6}}\right)\left(1-\frac{P_{\overline{NM}}}{10^{13}}\right),
\label{G}\\
\frac{dP_{R}}{dt}&=\left(\frac{100P_{N}k_{1}}{b}+P_{R}\left(\frac{1}{b_{R
}}-\frac{1}{d}\right)-
\frac{\left(3\times100+3\times10+1\right)P_{R}k_{1}}{b_{R}}\right)
\left(1-\frac{P_{\overline{NM}}}{10^{6}}\right)\left(1-\frac{P_{\overline{NM}}}{10^{13}}\right),
\label{R}\\
\end{align}
\end{subequations}
\begin{equation*}
\vdots
\end{equation*}

The equations for the other populations follow the same format and
can be derived from the state diagram and from the vector form of
the ODEs.  In words, Equation~\ref{D}, for example, says
 that the population of cells with a mutation in $D$ is increased by normal cells gaining a
  mutation in one of 100 genes in $D$ at a rate of $k_{1}$ every $b$ days.  The population is also
  increased by cells in state $D$ replicating (but not mutating) every $b$ days and dying
  every $d_{D}$ days.  The population is decreased by cells leaving state $D$ and gaining a single
  mutation in one of 3 other categories ($AD$, $DG$, $DR$) each with 100 genes,  by gaining a double
  mutation in one of 3 ways ($DGR$, $ADG$, $ADR$), with 10 genes being involved in each transition, or by gaining a triple mutation
  to go to state $ADGR$ with 1 gene being involved in the transition.  The logistic
  term caps the total population of non-normal, non-metastatic cells  at
  $10^{6}$ cells.  What is not visible in this standard form of the ODEs
  but is present in the vector form is the fact that the
  logistic angiogenesis cap is only imposed when less than 10\% of the non-normal, non-metastatic cells have a mutation in the $A$ category.
  Finally, populations of non-normal, non-metastatic cells are always capped by a lethal tumor
burden limit of $10^{13}$ cells.
\section{Calculation of cell division and cell death rates}
\label{cell div and cell death rates}
 In order to calculate the
change in cell division and cell death rates for mutations in $R$
and $D$, we begin with the assumption that birth and death rates
are equal for normal cells, i.e. cell division rate = $1/b$ = cell
death rate = $1/d$ = 1/10 days$^{-1}$.

We use an approximation to the ODEs where we treat the rate of
cells entering and leaving the state as negligible compared with
the tumor volume doubling time, since they are several orders of
magnitude different. Thus, for cells with mutations in $D$ but not
$R$, say, we consider
\begin{equation}
\frac{dD}{dt}=D\left(\frac{1}{b}-\frac{1}{d_{D}}\right).
\label{ln2exp1}
\end{equation}
Solving this gives $D = D_{0}
\exp\left(\left(1/b-1/d_{D}\right)t\right)$. A doubling
corresponds to $2=\exp\left(\left(1/b-1/d_{D}\right)T_{D}\right)$.
Similarly, for cells with mutations in $R$ and not $D$, we arrive
at $R = R_{0} \exp\left(\left(1/b_{R}-1/d\right)t\right)$. Taking
the natural logarithm of both sides leads to Equations~\ref{bR}
and~\ref{dD},
\begin{equation}
\frac{\ln 2}{1/b_{R}-1/d}=T_{R},
\label{bR}
\end{equation}
\begin{equation}
\frac{\ln 2}{1/b-1/d_{D}}=T_{D},
\label{dD}
\end{equation}
where $T_{R}$ is tumor volume doubling time for cells with a
mutation in $R$ but not $D$ and equals $T+50\times 3$, where
$T_{D}$ is the tumor volume doubling time for cells with a
mutation in $D$ but not $R$ and equals $T+50\times7$, and where
the $D:R$ importance ratio is $0.7:0.3$. The base tumor volume
doubling time is $T$ when the growing tumor has mutations in both
$D$ and $R$ (500 days). The value 50 is chosen to give realistic
doubling times for cells with mutations in $D$ (but not $R$) and
$R$ (but not $D$), on the upper bound of observed tumor volume
doubling times, where the cells typically have both mutations.
\section{Kinetics of various paths to cancer}
Given that multiple mutations are necessary to form a tumor, we
are interested in whether the specific order of mutations is
important. It is currently believed that the temporal sequence of
mutations determines the propensity of tumor
development~\citep{arends}. The extent to which genetic instability
($G$) determines the timing of tumorigenesis has been a
controversial issue in cancer biology.
 Some have argued that an increased premalignant mutation rate (that is, acquiring a mutation in $G$ early) is necessary for
 tumor development~\citep{loeb91,NRCvogelstein}.  Others have argued that an increased cell division rate, offering more opportunities to accumulate mutations,
 is sufficient for tumorigenesis~\citep{tomlinson2,tomlinson1,NRCtomlinson}.  Although the extent to which angiogenesis ($A$),
 decreased apoptosis ($D$), genetic instability ($G$), and increased replication rate ($R$) contribute to the development of cancer depends
 on the type of cancer involved, a better general understanding of the kinetics of various paths to cancer can be informative
  about their relative importance.

\begin{figure}[htpb]
\centering
  \subfigure[]{
  \includegraphics[width=8cm]{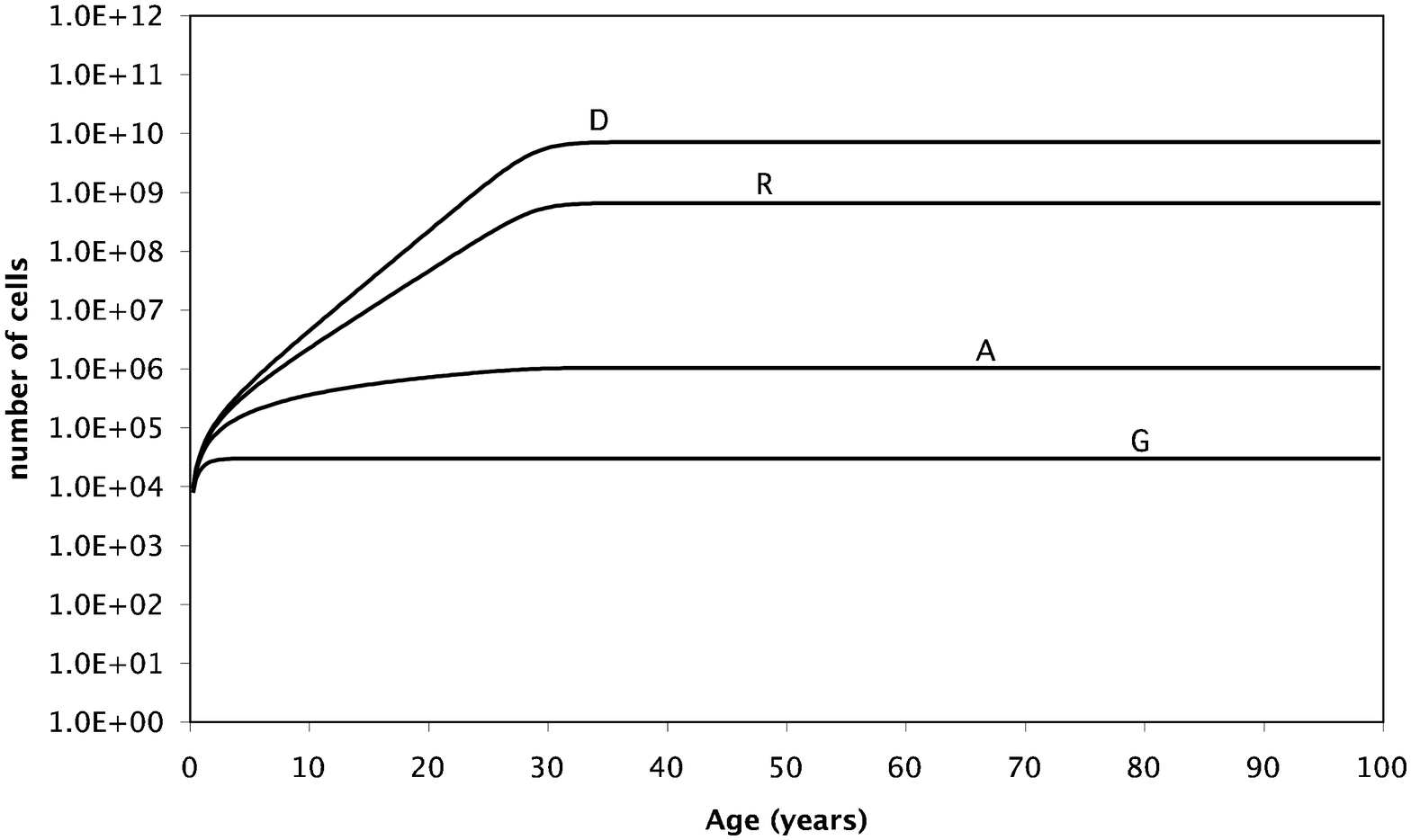}
  \label{odepath:a}
  }
    \subfigure[]{
  \includegraphics[width=7.5cm]{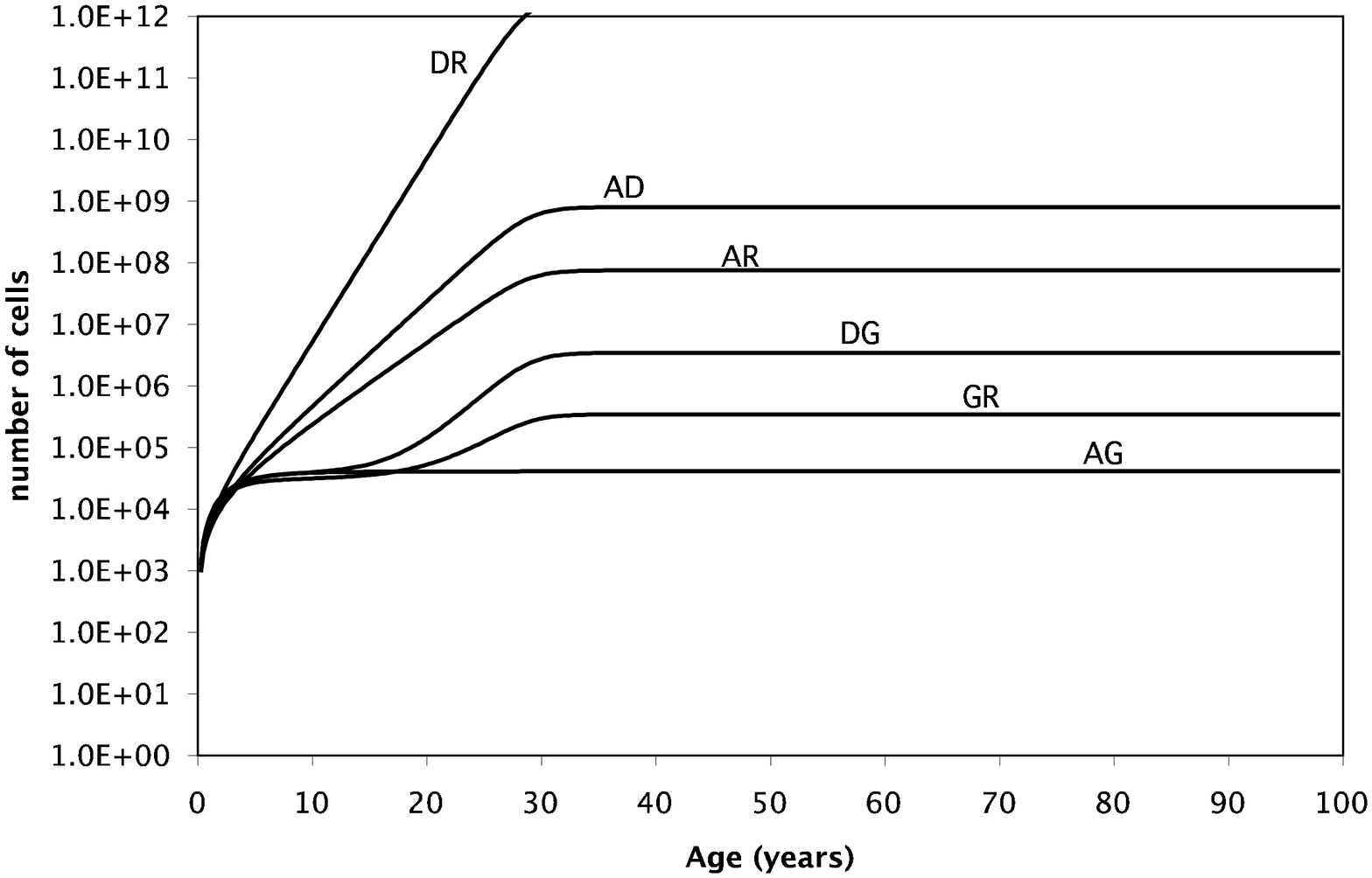}
  \label{odepath:b}
  }
  \subfigure[]{
  \includegraphics[width=8cm]{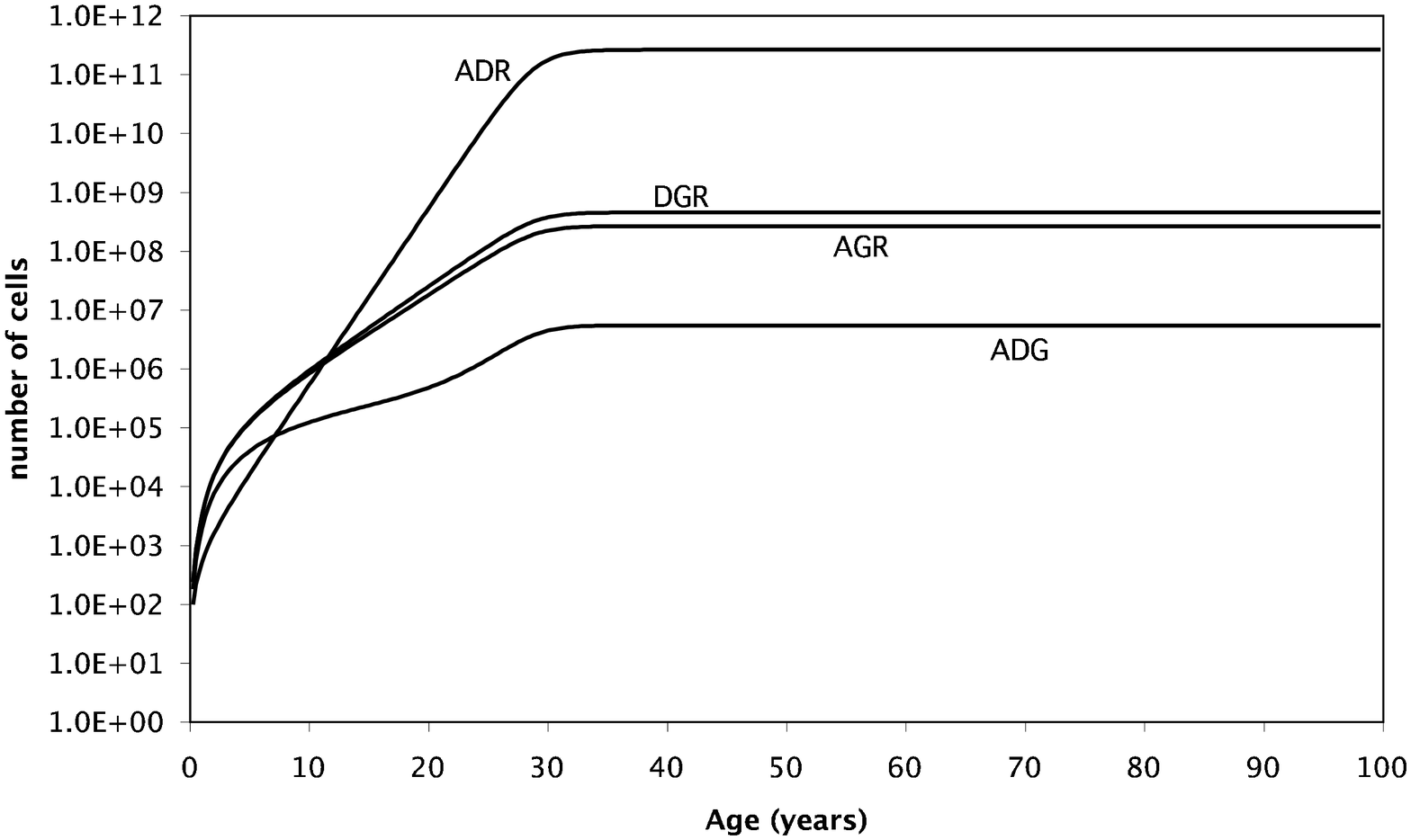}
  \label{odepath:c}
  }
  \subfigure[]{
  \includegraphics[width=8cm]{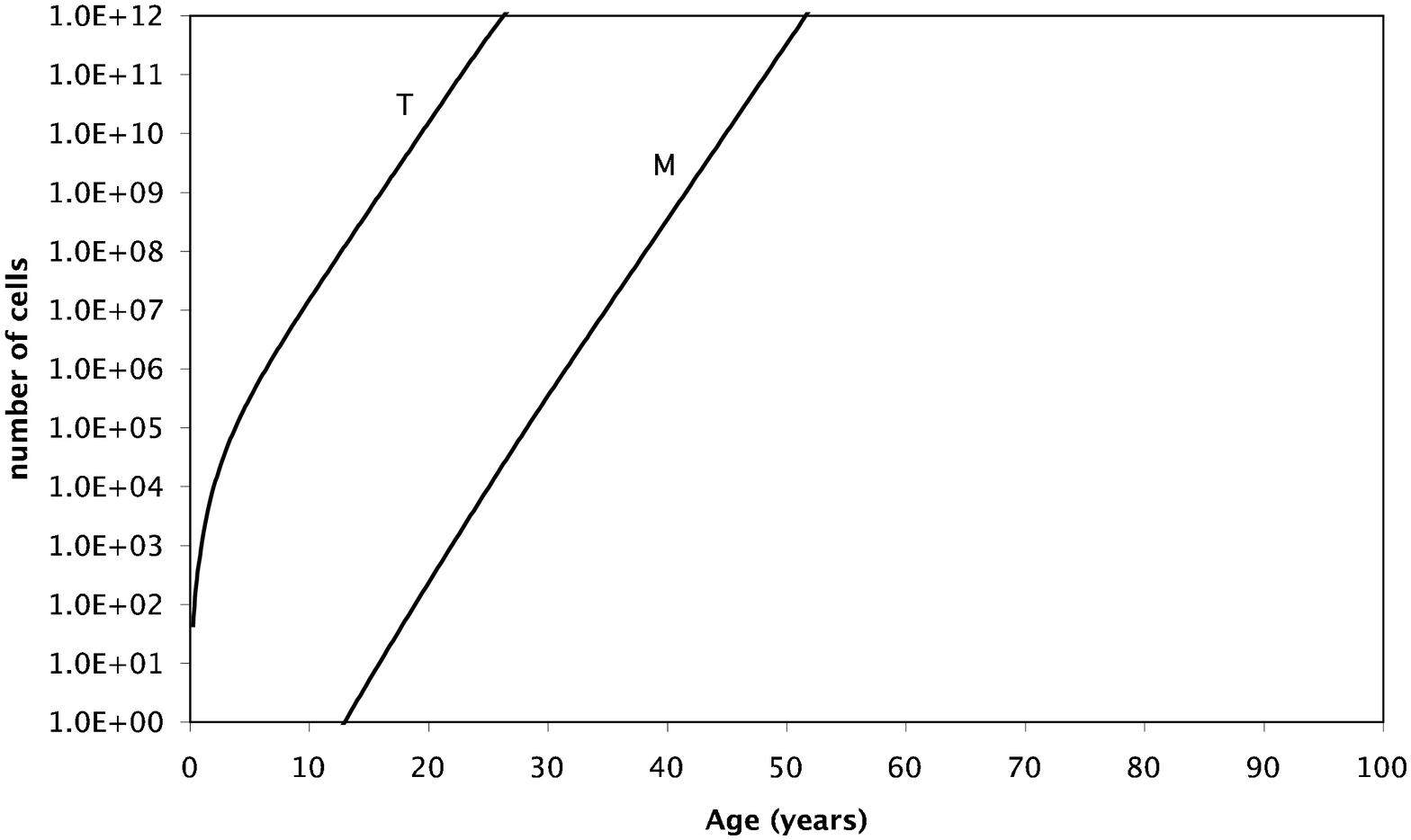}
  \label{odepath:tm}
  }
  \subfigure[]{
  \includegraphics[height=5.1cm]{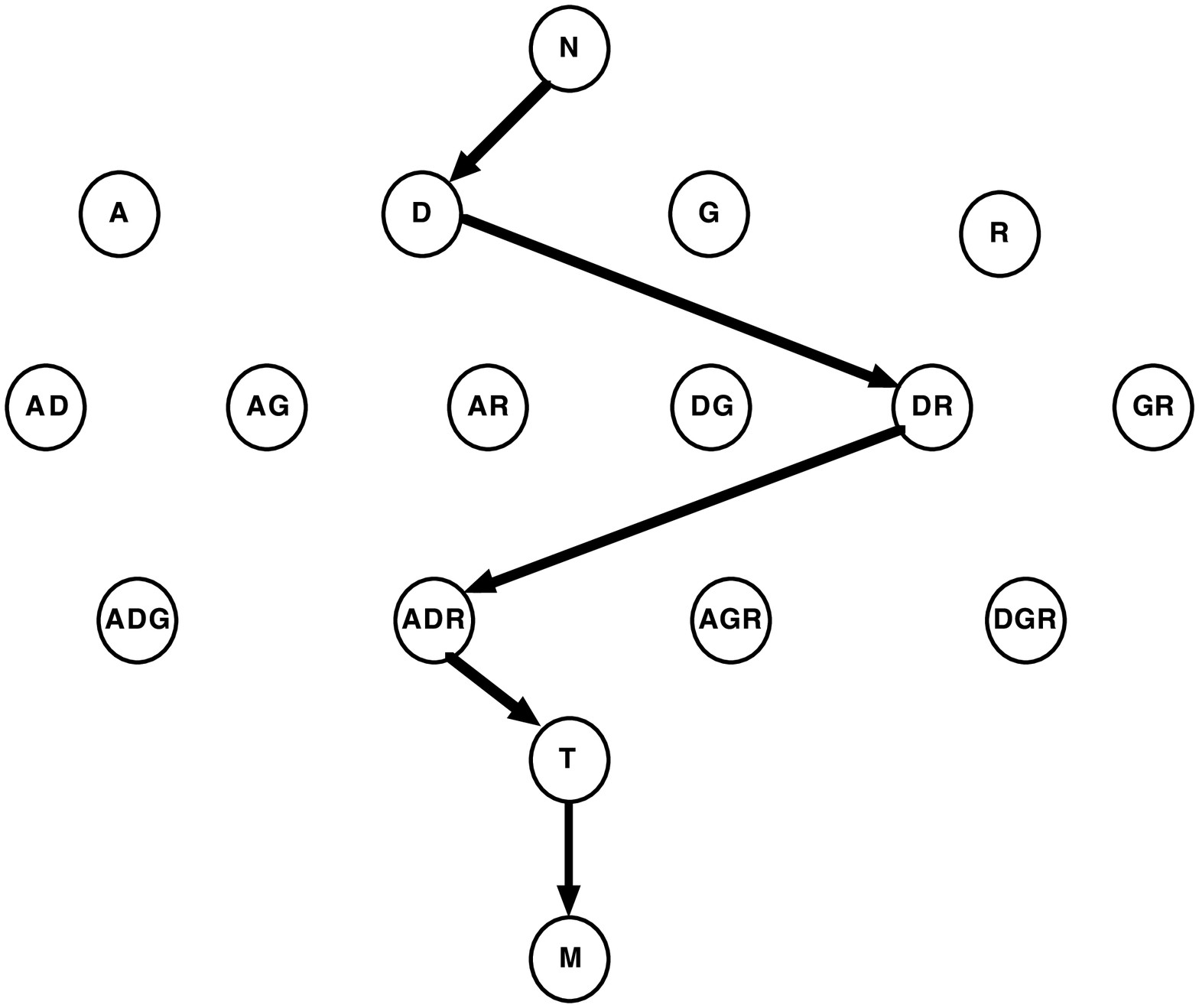}
  \label{odepath:d}
  }
\caption{Fastest path to cancer. (a) Dynamics of cell populations
with one type of mutation. (b) Dynamics of cell populations with
two types of mutations. (c) Dynamics of cell populations with
three types of mutations. (d) Dynamics of cell populations with
four types of mutations ($T$), and those that have metastasized
($M$). (e) The fastest path to cancer is by acquiring a mutation
in $D$, then $R$, then $A$, then $G$. } \label{odepath}
\end{figure}

We explore the kinetics of various pathways to cancer by analyzing
the dynamics of the different cell populations. By plotting
different sets of cell populations, we are able to identify the
individual contribution of each mutation to the development of
cancer. The growing populations of cells plateau at various points
in the graphs due to the imposed $10^{13}$ cell population cap. In
this model, the fastest pathway for tumor progression starts with
a mutation in $D$ (Figure~\ref{odepath:a}) which increases the
population of potential tumor cells. Next, a mutation in $R$ is
acquired, further increasing the population of cells by clonal
expansion (Figure~\ref{odepath:b}).  After acquiring these two
mutations, the tumor is sufficiently large to be inhibited by the
angiogenesis cap imposed by the model.  For this reason, a
mutation in the angiogenesis category occurs next in the fastest
path (Figure~\ref{odepath:c}). Finally, a mutation in  $G$
follows.  Figure~\ref{odepath:tm} shows
 the populations of tumor cells, $T$, and metastatic cells, $M$. Although the rate $k_{3}=10^{-9}$ is very low, the large increase
 in population of $T$ cells guarantees that eventually some cells successfully
 metastasize.

Our model predicts that genetic instability is more likely to be a
 feature of later-stage sporadic tumors, in accordance with the view of \cite{tomlinson1}. This is because
 a mutation in $G$ has no direct selective advantage, only an indirect advantage through increasing the mutation rates in other genes. Although genetic instability can
 aid tumorigenesis, selection and clonal expansion are the main driving force for tumor progression in this model, a
 conclusion which has been proposed previously by \cite{NRCtomlinson}.   
\section{Effect of inherited mutations on cancer development}
Here we examine the effect of different inherited mutations on
cancer development by varying our initial conditions.  Since most
inherited cancers are the result of mutations in tumor suppressors
~\citep{knudson3}, we model this situation by increasing the rate
of transition from a normal cell to the appropriate mutated cell
to $10^{-5}$ mutations/gene/cell division. This models a case
where a person inherits an inactivating mutation in one allele of
the gene. These cells are still functionally ``normal'' (thus they
begin in state $N$), but the chance of acquiring the second
``hit'' and losing functionality of the protein (moving into the
mutated state) is much increased.

\begin{figure}[phtb]
\centering
\includegraphics[height=7cm]{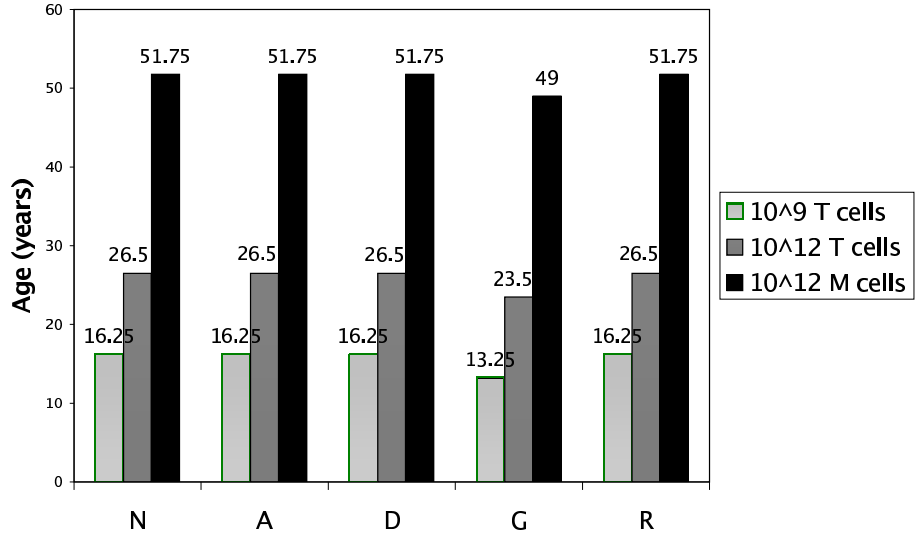}
\caption{Inherited mutations in cancer-critical genes.  Age at
which a person may acquire $10^{9}$ primary tumor cells, $10^{12}$
primary tumor cells, and $10^{12}$ metastatic cells with different
inherited mutations. For reference, the case where no mutations
are inherited is also shown ($N$).} \label{inborn}
\end{figure}

As expected, inheriting a mutation in a cancer-critical gene
decreases the time to cancer onset. The effects of inheriting a
mutation in each category on time to reach $10^{9}$ primary tumor
cells, $10^{12}$ primary tumor cells, and $10^{12}$ metastatic
cells are shown in Figure~\ref{inborn}. A tumor volume of 1 cubic
centimeter weighs about 1 gram and represents about $10^{9}$
cells~\citep{surgonc}. This tumor size is regarded as relatively
small in a clinical setting and it is at this size that a tumor
may give rise to the first symptoms and may first become
detectable by palpation~\citep{surgonc}. A tumor that weighs about
1 kilogram ($10^{12}$ cells) is approaching the lethal tumor
burden for a patient~\citep{surgonc}. The $10^{12}$ metastatic
cells plotted in Figure~\ref{inborn} are not necessarily localized
to one site in the body; they could represent $10^{12}$ cells
present in one location or $10^{11}$ cells present in each of 10
different locations, for example.

In contrast to the results obtained in 4.1.1 where the increased
population of cells caused by mutations in $D$ and $R$ dominates
the fastest path to sporadic cancer, {\it inheriting} a mutation
in a $G$ gene causes cancer onset at the earliest age.  There is
no observable difference between inheriting a mutation in one of
the other categories and inheriting no mutations at all.   In the
fastest path plots (Figure~\ref{odepath}), there is equal
probability of acquiring a mutation in $A$, $D$, $G$ or $R$. $D$
will dominate over $G$ due to the fact that the transition from
one state to another is a function not only of the mutation rates
$k_{1}$ and $k_{2}$ but also the cell population size. Both $D$
and $G$ are equally likely to begin with, but since $D$ increases
the net cell population very quickly, it soon dominates over the
rate $k_{2}$ associated with $G$. Therefore, the fastest path to
sporadic cancer is through a mutation in $D$ first.

In comparison, when a mutation in $G$ is {\it inherited}, the cell
has already surpassed the initial probability hurdle of acquiring
a mutation in $G$.
 The rate of subsequent mutation is now 1000-fold higher and once a mutation in $D$ or $R$ is obtained,
 the cell population will begin to increase.
 For this reason, an inherited mutation in $G$ has the greatest
 effect. This is consistent with the fact that many inherited
 cancer syndromes are the result of a mutation in the $G$ category.  These include
 xeroderma pigmentosum, ataxia telangiectasia, Nijmegen breakage
 syndrome, hereditary non-polyposis colorectal cancer, and Bloom
 syndrome~\citep{NRCtomlinson}.

 The time to develop a palpable primary tumor ($10^{9}$ cells) in
 this model is 16.25 years if no mutations are inherited (Figure~\ref{inborn}).  Even taking into account
 the fact that detection of the tumor would not occur until several years
 later~\citep{surgonc}, this age of cancer onset is significantly earlier than the
 average age of cancer onset in the human population~\citep{ageofcancer}.  This is an indication of the need for more accurate information on
 cell division, cell death, and tumor doubling rates. Importantly, this may also be an indication
 that acquisition of mutations in more than four categories is
 necessary for development of a primary tumor.  Adding two more steps to the multistep model would certainly delay the time to
 cancer, and would be more in accordance with the six-step model
 proposed by \cite{hanahan00}.  Consideration of the role of the
immune system in curbing the growth of a tumor would also slow the
time to cancer onset.  Our model does not directly consider this
factor, although category $D$ does allow for apoptosis initiated
by the immune system.  Consideration of these three factors would
allow the model to be more appropriately scaled to the timing of
human cancer.
\section{Sensitivity analysis of variations in the parameters}
In order to determine the relative contributions of the parameters
to the model, we vary each parameter listed in
Table~\ref{paramtable} while holding all others constant at the
default value.  The default values chosen are our best estimate
from the literature.  Except for {\it $D$:$R$ importance ratio}
where we use 0.3:0.7 to determine the effect on the fastest path,
and {\it \% $A$ cells needed to remove cap} where we test the
range from 0\% to 100\%, the other values are chosen to be near
the upper and lower bounds of the range given in the literature.
We assess the contribution of each by examining their effect on
time to reach $10^{12}$ $M$ cells and on the fastest pathway to
cancer in Table~\ref{paramtable}.

The most salient result of the sensitivity analysis is the
robustness of the model.  Despite trials with very high values for
$k_{2}$, the fastest path to somatic cancer is always via a
mutation in $D$ then $R$ then $A$ then $G$, except in the case
where we flip the $D$:$R$ importance ratio.  As expected, a ratio
of 0.3:0.7 flips the roles of $D$ and $R$ in the fastest path to
give $RDAG$, but does not change the time to $10^{12}$ $M$ cells
from the default value of 51.75 years. Using a $D$:$R$ ratio of
0.8:0.2 decreases the time to reach $10^{12}$ $M$ cells due to the
increased weight given to $D$.

\begin{figure}[pbth]
\subfigure[]{
  \includegraphics[width=8cm]{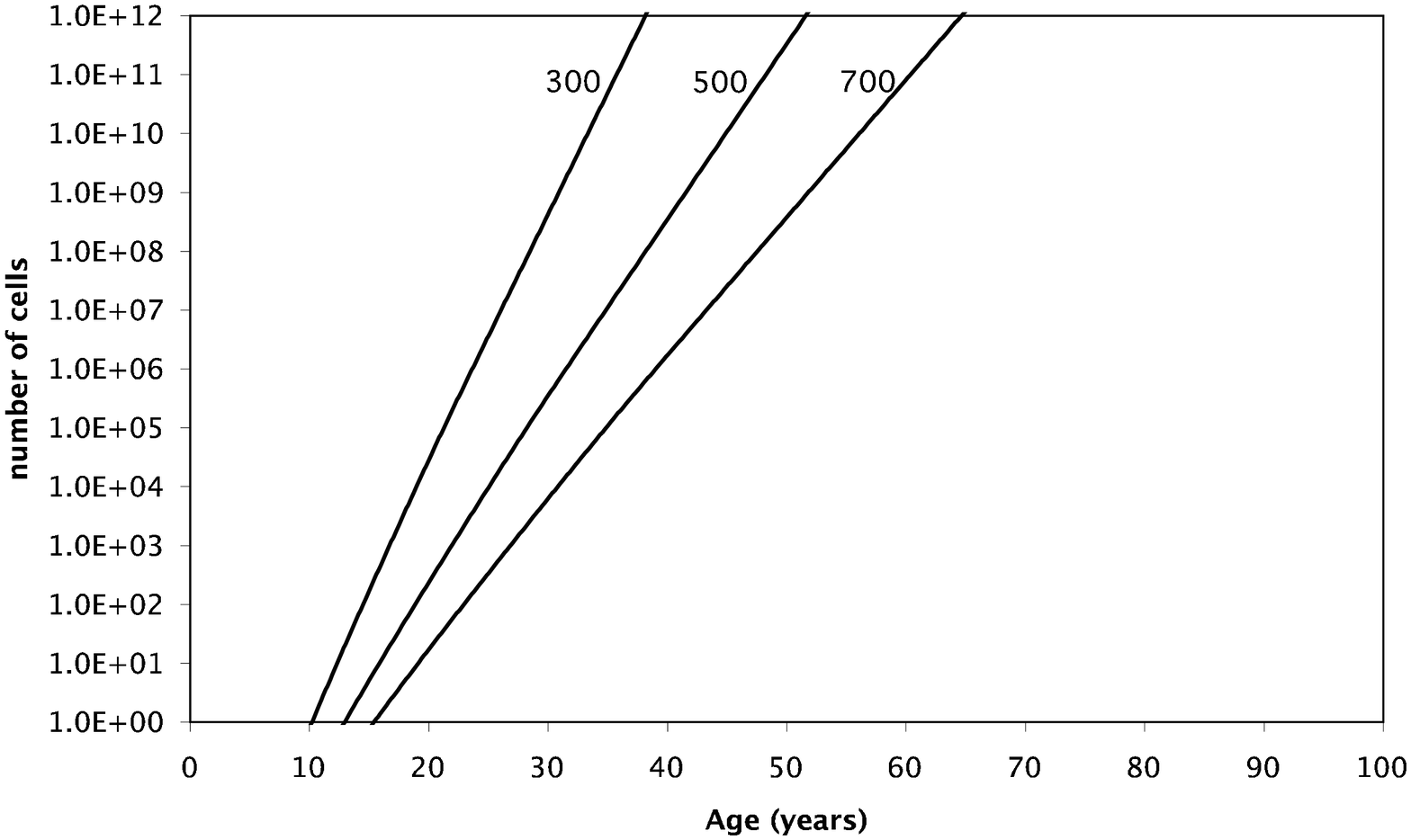}
  \label{tvdt}
  }
\subfigure[]{
  \includegraphics[width=8cm]{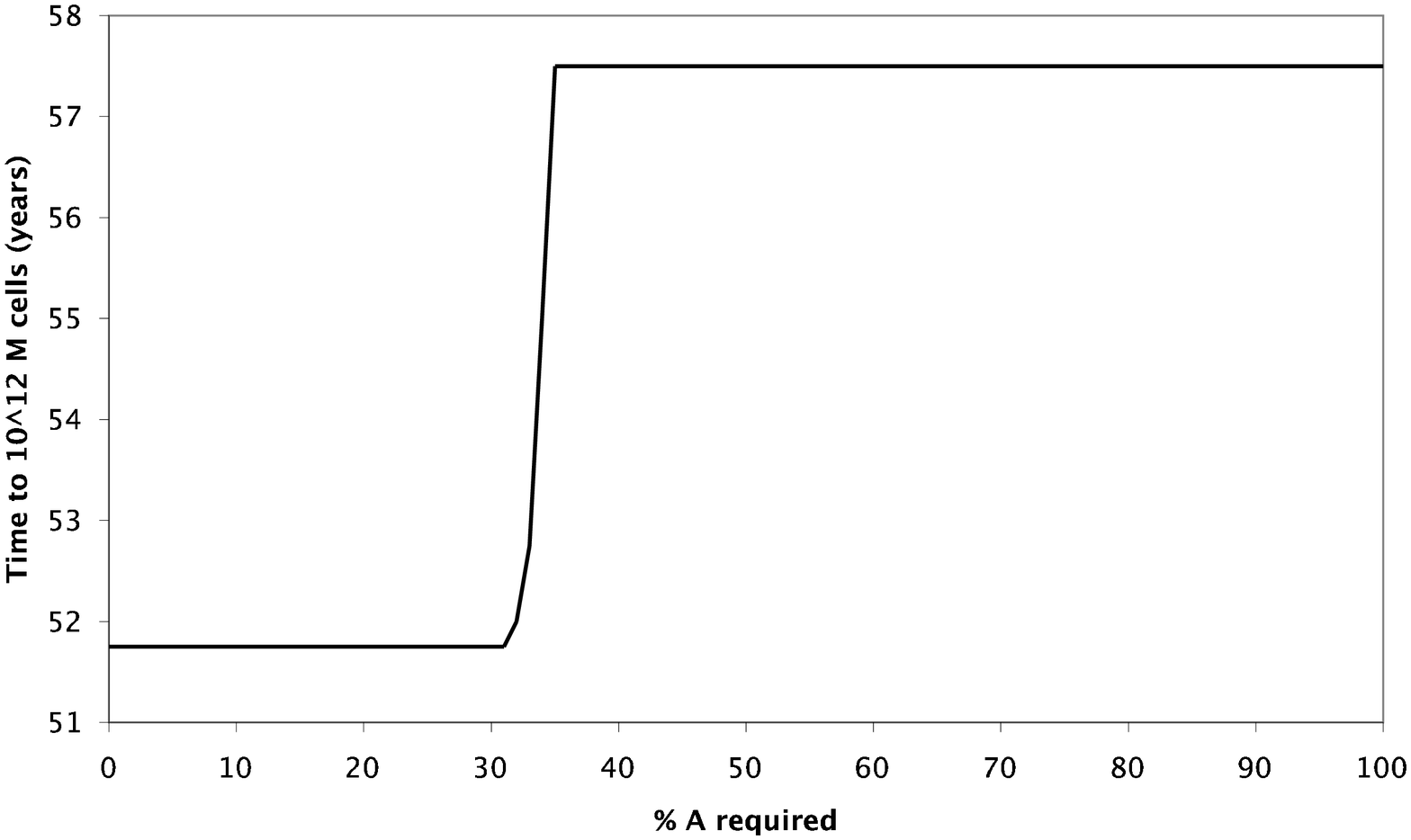}
  \label{varA}
  }
\caption{Sensitivity of the model to changes in parameters.  (a)
Sensitivity to changes in tumor volume doubling time. Time to
reach $10^{12}$ $M$ cells is shown for tumor volume doubling times
of 300 days, 500 days, and 700 days. (b) Effect of variations in
percentage of $A$ cells required to induce angiogenesis on time to
reach $10^{12}$ $M$ cells.} \label{sensanal}
\end{figure}

The parameter that has the largest effect on time to reach
$10^{12}$ $M$ cells is the tumor volume doubling time.  A tumor
volume doubling time of 300 days decreases the time to reach
$10^{12}$ $M$ cells by 13.50 years relative to the default of 500
days, and a tumor volume doubling time of 700 days increases the
time to reach $10^{12}$ $M$ cells by 13.00 years.  This effect is
seen in Figure~\ref{tvdt} as well as in Table~\ref{paramtable}.
The large effect of this parameter on the model is due to its
impact on the $\left(\frac{1}{b}-\frac{1}{d}\right)$ term; when
cells have mutations in $R$ and/or $D$, the terms become $1/b_{R}$
and/or $1/d_{D}$, allowing the cell populations to increase at a
rate that reflects the tumor volume doubling time chosen.

Variations in the birth and death rates to 1 every 5 days and 1
every 30 days also have an effect on time to reach $10^{12}$ $M$
cells. This can be seen in row one of Table~\ref{paramtable}, but
the effect is small when compared with the effect of tumor volume
doubling time.

The time (51.75 years) to reach $10^{12}$ $M$ cells does not
change in varying the percentage of $A$ cells needed to remove the
angiogenesis cap from 0\% to 31\%.  Between 31\% and 35\%, the
time to reach $10^{12}$ $M$ cells increases rapidly.  The time
(57.50 years) to reach $10^{12}$ $M$ cells does not change in
varying the percentage from
 35\% to 100\%.  This effect can be seen in Figure~\ref{varA}.  At 31\%, the requirement
 for mutations in the $A$ category begins to have an effect on the
 growing cell populations.  At 35\% and above, the percentage of $A$ cells
 required is so large that sum of the populations of non-normal,
 non-metastatic cells never gets above $10^{6}$ because there are never at least 35\% with $A$
 mutations.  Thus the time to reach $10^{12}$ $M$ cells depends only on a fixed number of
$T$ cells in each case, and remains constant at 57.50 years for
percentages 35\% and above.

A ten-fold change (from $10^{-7}$ to $10^{-6}$ mutations/gene/cell
division) in mutation rate without a $G$ mutation ($k_{1}$) has a
larger effect on time to reach $10^{12}$ $M$ cells than a ten-fold
change (from $10^{-4}$ to $10^{-3}$ mutations/gene/cell division)
in mutation rate with a $G$ mutation ($k_{2}$). This is due to the
fact that the effect of $k_{2}$ only becomes important later in
tumorigenesis since $G$ is last in the fastest path to cancer,
whereas the effect of $k_{1}$ occurs at the beginning. There is no
change in time to reach $10^{12}$ $M$ cells when $k_{2}$ is
increased from $10^{-3}$ to $10^{-2}$ mutations/gene/cell division
because the effect of mutation rate has already saturated the
system at a $k_{2}$ value of $10^{-3}$.

Increasing the number of genes involved in the transitions
decreases time to reach $10^{12}$ $M$ cells simply due to the
larger numerator in the differential equations.

\begin{table}[htb]
\scriptsize \centering \caption{Sensitivity of the model to
changes in changes in parameters.  Cell birth and death rates have
units days$^{-1}$. Tumor volume doubling times are measured in
days. Mutation rates are measured as mutations/gene/cell division.
Number of genes involved in transitions are listed as number
involved in single, double, triple transitions.  ``Other'' refers
to different values tested in the sensitivity analysis.  ``Time''
refers to age at acquisition of $10^{12}$ $M$ cells for a
variation in that parameter, measured in years. ``Path'' refers to
the fastest path to cancer for a variation in that parameter.  }
\hspace*{\fill}\newline
\begin{tabular}{|c||c|c|c||c|c|c||c|c|c|}
\hline Parameter & Default & Time & Path & Other & Time &
Path & Other  & Time & Path
\\\hline\hline Cell birth and death rates & 1/10 &  51.75 & DRAG & 1/5 & 50.25 & DRAG & 1/30 & 54.75 & DRAG
\\\hline Tumor volume doubling time & 500 &  51.75 & DRAG & 300 & 38.25 & DRAG & 700 & 64.75 & DRAG
\\\hline \% $A$ cells needed to remove cap & 10\% &  51.75 & DRAG & 30\% & 51.75 & DRAG & 40\% & 57.50 & DRAG
\\\hline $D:R$ importance ratio & 0.7:0.3 &  51.75 & DRAG & 0.8:0.2 & 50.50 & DRAG & 0.3:0.7 & 51.75 & RDAG
\\\hline Mut. rate with a $G$ mutation & $10^{-4}$ &  51.75 & DRAG & $10^{-3}$ & 50.50 & DRAG & $10^{-2}$ & 50.50 & DRAG
\\\hline Mut. rate without a $G$ mutation & $10^{-7}$ &  51.75 & DRAG & $10^{-6}$ & 48.50 & DRAG &  &  &
\\\hline \# of genes involved in transitions & 100, 10, 1 &  51.75 & DRAG & 500, 100, 10 & 48.25 & DRAG &  &  &
\\\hline
\end{tabular}
\label{paramtable}
\end{table}

\section{Discussion}
This paper explores facets of the multistep model of oncogenesis.
The key findings of this paper are (1) the fastest path to somatic
cancer is predicted to be through gaining mutations in $D$, then
$R$, then $A$, then $G$, (2) of the four categories of mutations,
inheriting a mutation in $G$ is predicted to produce cancer at the
earliest age, and (3) the fastest path to somatic cancer is robust
to realistic changes in parameters, with the model being most
affected by variations in tumor volume doubling time.

The strength of our model lies not in its utility for predicting
any one individual's time to cancer onset {\it per se}, but rather
in the fact that it presents a novel approach to understanding the
genetic basis of cancer from a systems biology perspective.
Although a thorough testing of this model is not currently
possible due to lack of appropriate biological data, this model
establishes the groundwork for future models that can be directly
tied to clinical and molecular data. We hope that the creation of
this model for the multistep progression to cancer will encourage
biologists to gather quantitative data and will suggest which
experiments should be performed with highest priority. The only
parameter values which are reasonably agreed upon in the
literature are the spontaneous mutation rate and the size to which
a tumor can grow before angiogenesis is required. All other
parameter values could use experimental refinement, especially
tumor volume doubling time, since it has the largest effect on
time to cancer onset. Better estimates of parameter values,
inclusion of two additional categories to give a total of six
steps in the multistep model, and consideration of the role of the
immune system in curbing the growth of a tumor will allow our
model to be more appropriately scaled to human cancers. Modeling
the multistep accumulation of genetic mutations in cancer will
give insight into topical questions about the progression of a
normal cell to a cancerous cell, enabling cancer treatments to be
better targeted to various stages of cancer progression, and
suggesting the most important directions for future experimental
research.
\section{Acknowledgements}
We thank Trachette Jackson, Gilbert Omenn, 
John Holland (University of Michigan),
Andrew Allison, Samuel Mickan, David Findlay, Brendon Coventry (The University of
Adelaide), Setayesh Behin-Ain (The Queen Elizabeth Hospital,
Australia),  and Roger Reddel
(Children's Medical Research Institute, Australia) for useful
discussions, Sara Dempster (Massachusetts Institute of Technology)
for her early contributions to this project, and the Santa Fe
Institute for the opportunity to attend the Complex Systems Summer
School where we began our exploration of this topic.  This work
was supported in part by scholarship funding from The Queen
Elizabeth Hospital, Adelaide, Australia, arranged through Setayesh
Behin-Ain.
\bibliography{phd}
\bibliographystyle{elsart-harv}
\end{document}
\end